 \documentclass[5p,times]{elsarticle}

\usepackage{amssymb}
\usepackage{lipsum}

\usepackage{amsmath,amssymb,amsfonts}%
\usepackage{graphicx}
\usepackage{dcolumn}
\usepackage{bm}
\usepackage{derivative}
\usepackage{xcolor}
\usepackage{scrextend}
\usepackage{graphicx}
\usepackage{url}
\usepackage{mathtools, cuted}
\usepackage{hyperref}

\DeclareMathOperator\erfc{erfc}

\journal{Physics Letters B}

\begin{document}

\begin{frontmatter}



\title{Quantum Big-Bounce as a phenomenology of RQM in the Mini-superspace}


\author[first]{Simone Lo Franco}
\ead{lofranco.1843692@studenti.uniroma1.it}
\author[second,first]{Giovanni Montani}
\ead{giovanni.montani@enea.it}
\affiliation[first]{organization={Department of Physics, ``Sapienza'' University of Rome},
            addressline={P.le Aldo Moro, 5}, 
            city={Roma},
            postcode={00185}, 
            country={Italy}}
\affiliation[second]{organization={ENEA, Fusion and Nuclear Safety Department, C.R. Frascati},
            addressline={Via E. Fermi, 45},
            city={Frascati},
            postcode={00044},
            state={Roma},
            country={Italy}}

\begin{abstract}
We investigate the emergence of a quantum Big-Bounce in the context of an isotropic Universe, filled by a self-interacting scalar field, which plays the role of a physical clock. The bouncing cosmology is the result of a scattering process, driven by the scalar field potential, which presence breaks down the frequency	separation of the Wheeler-DeWitt equation, treated in strict analogy to a relativistic quantum system. Differently from previous analyses, we consider a really perturbative	self-interaction potential, affecting the dynamics in a finite range of the time labeled by the scalar clock (and in particular we remove the divergent character previously allowed). The main result of the present analysis is that, when the Relativistic Quantum Mechanics formalism is properly implemented in the Mini-superspace analogy, the probability amplitude for the bounce is, both	in the standard and polymerized case, characterized by a maximum in	correspondence of the quasi-classical condition of a Universe minimum volume.
\end{abstract}




\end{frontmatter}




\section{Introduction}
\label{Introduction}
One of the most important successes of the Loop Quantum Gravity \cite{Thiemann_2007,Montani:CanQG} has been the proof that the geometrical area and volume operators kinematically possess a discrete spectrum \cite{Rovelli_1995}. A significant implication of this result has been found in the cosmological implementation of the general theory, known as Loop Quantum Cosmology \cite{Ashtekar_2011_review}. In fact, despite some limitations emerging in the cosmological formulation \cite{CIANFRANI_2012,Cianfrani_2012_2,Bojowald_2019}, the emergence of a Big-Bounce configuration is certainly an intriguing achievement \cite{Ashtekar_2006_quantumBB,Ashtekar_2006_quantumBB_2} (for recent developments in this field, which enforce the original analysis, see \cite{Bruno_2023,Bruno_2023_2}). However, the Big-Bounce picture emerging in Loop Quantum Cosmology is essentially a semiclassical one, and that this	situation can have a general nature is not assessed \cite{Bojowald_2020}.  As soon as a general enough model is considered, see for instance the Bianchi Universes \cite{Montani:PrimCos,Bojowald_2004}, the possibility to construct localized states, approaching the singularity, is significantly limited by the behavior of the Universe anisotropy. Thus, in \cite{Giovannetti_2022_bianchi} was first proposed the idea that a Big-Bounce can take place for a Bianchi I model as the result of a scattering process, on a pure quantum domain.	The basic statement was the identification of the Wheeler-Dewitt equation for the model with a Klein-Gordon theory, which time corresponded to the isotropic Minser variable and which spatial coordinates were identified in corresponding anisotropic variables \cite{thorne2000gravitation,misner1968isotropy,misner_69}. The possibility for such an identification has been investigated in detail in \cite{Wald_1993,Higuchi_1995}, but in \cite{Giovannetti_2022_bianchi} the introduction of an ekpyrotic-like time-dependent term \cite{Lehners_2008} allowed for a non-zero transition amplitude from negative to positive frequency solution, i.e. a quantum Bounce configuration has been inferred. The analysis in question was then extended to the employment of a self-interacting scalar field as a relation clock for the isotropic Universe \cite{Giovannetti_2023}. The results emerging from these two analyses consist of dealing with a maximum probability amplitude for the Big-Bounce, when the mean momentum of the incoming packet is equal to the mean value of the outcoming momentum. In this paper, we consider a similar analysis to the one in \cite{Giovannetti_2023}, trying to address an important question, that remained open in the previous studies: the divergence of the scattering (time-dependent) potential in one direction of the time axis. Despite in \cite{Wald_1993}, it was argued that such a feature does not prevent the introduction of a suitable Hilbert space, nonetheless, questions arise on the real viability of the relativistic quantum scattering, especially because of possible (hidden) violation of the unitarity of the S-matrix \cite{Bjorken:Drell}. Here, we consider a potential term for the scalar field which the ekpyrotic models inspire \cite{Steinhardt_2002_ekp_pot1,Steinhardt_2002_ekp_pot2,Buchbinder_2007_new_ekp}, that has the important feature to be significantly non-zero only in a finite domain (and its amplitude there is controlled by a coupling constant). Then, we repeat all the analysis of the relativistic quantum scattering (in the absence of new degrees of freedom creation, or their annihilation), both for the standard Wheeler-Dewitt states and for the case in which Polymer Quantum Mechanics \cite{Corichi_2007,Corichi_2007_2,Barca_2021,Mandini} is implemented. The relevant achievement of the present analysis consists of demonstrating that, if the perturbative character of the scattering potential is guaranteed, then the quantum Big-Bounce is most probable when the quasi-classical condition of a minimal Universe volume is fulfilled. Furthermore, we show that, as soon as, the potential is non-perturbative, the transition amplitude acquires the same features discussed in \cite{Giovannetti_2023}. This suggests that the perturbative character of the scattering potential is a basic requirement to implement the equipment of the Relativistic Quantum Mechanics scattering process. It remains however to be noted that, differently from here, in the two previous analyses the localized states were constructed through exact solutions of the quantum theory, including the perturbation potential. This fact suggests a possible reformulation of the analyses with diverging potential in a different scenario, avoiding the direct use of the Feynman propagator \cite{Feynman}.

\section{\label{classical} Classical and quantum description of the closed FLRW model}
	In this section, we study both the classical and quantum evolution of a closed FLRW model. The ADM formalism is selected to describe the classical behavior of such a Universe, while the Wheeler-DeWitt (WDW) theory will be used to move to a quantum mechanical framework. Furthermore, we show how introducing a cutoff in the flat FLRW can reproduce the evolution of the closed one near the cosmological singularity.
	\subsection{Classical framework}%
	A closed isotropic Universe, filled with a self-interacting massless scalar field $\phi(t)$, is described by the Hamiltonian constraint%
	\begin{equation}%
		\frac{NV}{C} e^{-3\alpha} \left[ -p^2_\alpha + \frac{6}{\chi} p_\phi^2 - V_K(\alpha) +V_S(\phi, \alpha) \right] = 0\,,
		\label{eq-ham-constr}%
	\end{equation}%
	where $N$ is the lapse function, $\alpha = \ln{(a)}$ is the isotropic Misner variable, $p_\phi\,,p_\alpha$ are the conjugated momenta, $V=2\pi^2$ is the  fiducial volume, $\chi = 8\pi G/c^4$ and $C=12V^2/\chi$. The potential term $V_K(\alpha) = (3 C K / \chi) e^{4\alpha}$ arises from the presence of positive spatial curvature $K$, while the self-interaction potential $U(\phi)$ is responsible for the term $V_S(\phi, \alpha) = C e^{6\alpha} U(\phi)$. The classical evolution of the system can be derived from the set of Hamilton equations
	\begin{subequations}%
		\begin{equation}%
			\dot{\alpha} = -\frac{\chi N}{6V} p_\alpha e^{-3\alpha}\,, \qquad \dot{p}_\alpha = \frac{\chi N}{V}\left(\frac{V_K}{3}-\frac{V_S}{2}\right)\,,%
			\label{eq-ham-eqs1}
		\end{equation}%
		\begin{equation}%
			\dot{\phi} = \frac{N}{V} p_\phi e^{-3\alpha}\,, \qquad \dot{p}_\phi = -NVe^{3\alpha}\pdv{U(\phi)}{\phi}\,.%
			\label{eq-ham-eqs2}
		\end{equation}%
		\label{eq-ham-eqs}
	\end{subequations}%
	Since we will consider $V_S(\alpha, \phi)$ as a scattering potential we are interested in the limit where $U(\phi)$ is negligible. In such a regime, $p_\phi$ is a constant of the motion. To use $\phi$ as the internal time of the system \cite{rovelli_internal_time} it is interesting to derive the evolution law $\alpha_K(\phi)$, where the subscript $K$ denotes that the trajectory is related to the closed FLRW model. This quantity can be derived by combining the Hamilton equations (\ref{eq-ham-eqs}) for $\alpha$ and $\phi$, finding the differential equation $d\alpha/d\phi = -p_\alpha/p_\phi$. Using the scalar constraint in Eq. (\ref{eq-ham-constr}) and setting $3 C K/ \chi = 1$ to lighten up the notation, this equation yields the evolution law \footnote{We consider the rescaled field $\phi \to \sqrt{\chi/6}\, \phi$.}
	\begin{equation}
		\alpha_K(\phi) = \frac{1}{4} \ln{\left[p_\phi^2 - p_\phi^2 \tanh{(2\phi)} \right]}\,.%
		\label{eq-classial-closed}
	\end{equation} 
	As expected, the Universe is subjected to an expansion phase from the cosmological singularity placed at $\alpha \to -\infty$ up to the turning point placed at $\alpha_{tp}=(1/2)\ln{(p_\phi)}$. Then, the Universe contracts towards the cosmological singularity. The scalar field $\phi$ can, indeed, represent a good candidate as the internal time of the system, since it follows monotonically the evolution with respect to time when $N = V e^{3\alpha}/p_\phi$. We now compare the effects of the curvature potential on the dynamics with the imposition of a cutoff on the evolution of a flat isotropic Universe. For a flat FLRW model, $V_K(\alpha) \equiv 0$, hence, the expansion and contraction trajectories of $\alpha$ with respect to the internal time $\phi$ are given by $\alpha_\pm (\phi) = \pm \phi + c_\pm$, where $c_\pm$ are integration constants. By imposing the condition $\alpha_0(\phi) \leq \alpha_{\text{max}}$ and properly linking the two phases, we obtain the evolution law
	\begin{equation}%
		\alpha_0(\phi) = - |\phi| + \alpha_{\text{max}}\,.%
		\label{eq-classical-flat}
	\end{equation}%
	Here the subscript $0$ denotes that we are considering the flat $K=0$ case. By fixing a proper value of $\alpha_{\text{max}}$, this evolution law reproduces the behavior of the closed Universe near the singularity and the presence of a turning point in the dynamics.
	
	\subsection{Quantum framework}
	The application of the Dirac quantization scheme to the ADM formalism leads to the WDW theory \cite{Montani:PrimCos,Montani:CanQG}, where the constraints are now promoted to quantum operators acting on some Hilbert space. Hence, the super-Hamiltonian constraint leads to the WDW equation, which selects the state allowed from the theory. In the case of a closed FLRW model, the scalar constraint in Eq. (\ref{eq-ham-constr}) yields the WDW equation, in Planck units $\hbar = c = G = 1$,
	\begin{equation}%
		\left[ \partial_\alpha^2 - \partial_\phi^2 - V_K(\alpha) + V_S(\phi, \alpha) \right] \Psi(\phi, \alpha) = 0\,,
		\label{eq-WDW-complete}
	\end{equation}%
	where $\Psi(\phi, \alpha)$ is the Universe wave function. Such an equation is analogous to a Klein-Gordon one. This analogy can be exploited to avoid the {\it problem of time} that rises in the framework of Canonical Quantum Gravity. In the spirit of a {\it time after quantization} approach, the scalar field $\phi$ in Eq. (\ref{eq-WDW-complete}) can be regarded as the internal time of the system. In this sense, the self-interaction potential $U(\phi)$ is a time-dependent potential, and its contribution to the WDW equation $V_S(\phi, \alpha)$ can be treated as a scattering potential in the framework of Relativistic Quantum Mechanics. Hence, Eq. (\ref{eq-WDW-complete}) can be equipped with the KG-like inner product 
	\begin{equation}
		\left(\psi(\phi,\alpha)\,,\, \varphi(\phi, \alpha) \right) = i \int_{-\infty}^{+\infty} d\alpha\ \psi^*(\phi, \alpha) \overleftrightarrow{\partial_\phi} \varphi(\phi,\alpha)\,.%
		\label{eq-KG-inner}
	\end{equation}    
	Such an inner product can be used to construct probability densities and study the semiclassical evolution of the Universe wave function. However, these probabilities are positive-definite only for pure positive/negative solutions of Eq. (\ref{eq-WDW-complete}). Nonetheless, Eq. (\ref{eq-KG-inner}) can be used to construct the scattering propagator theory of RQM \cite{Bjorken:Drell,weinberg_1995}. Thus, we will describe the bounce as a quantum transition mediated by the contribution from the self-interaction potential $U(\phi)$, in the spirit of \cite{Giovannetti_2022_bianchi, Giovannetti_2023}. To ensure the unitarity of the scattering operator, the self-interaction potential must vanish at very early/late times. The set of asymptotic states of the process is the set of orthonormal solutions of Eq. (\ref{eq-WDW-complete}) in the limit $U(\phi) \to 0$. These solutions, which have been found in \cite{kiefer-curv-sol,de_Cesare_2016} and are frequency-separated, take the form
    \begin{equation}
        \chi^\pm_k (\phi,\alpha) = \sqrt{\frac{\sinh{(\pi k/2)}}{2 \pi^2}}\,  e^{\pm i k \phi}\, K_{\frac{ik}{2}}\left( \frac{1}{2}e^{2\alpha} \right)\,, \label{curve:wdw:bas:sol}    
    \end{equation}
    where $K_\nu (x)$ are the Macdonald functions and $k\leq 0$. Exploiting the orthogonality of the Macdonald functions \cite{PASSIAN2009380_macdonald,szmytkowski2009comment_macdonald}, it can be shown that such solutions satisfy the relations 
	\begin{equation}
		\left(\chi^\pm_{k'} \,,\,  \chi^\pm_k\right)= \pm \delta(k-k')\,, \qquad \left(\chi^\mp_{k'} \,,\, \chi^\pm_k\right)=0\,.%
		\label{eq-orthorel-macd}
	\end{equation}
	Localized wave packets, constructed from positive (negative) frequency solutions follow the semiclassical trajectory. It has been shown in \cite{de_Cesare_2016} that it is possible to construct a Feynman propagator for this equation through a Mini-superspace variables transformation, exploiting the reparametrization invariance of the theory. However, to keep a straightforward interpretation of the Mini-superspace variables when studying the bounce as a pure quantum mechanical process, we prefer not to use this strategy. Instead, to construct a meaningful propagator theory, we replace the curvature potential $V_K(\alpha)$ with the cutoff boundary condition
	\begin{equation}
		\Psi(\phi,\alpha \geq \alpha_{\text{max}}) = 0\,,%
		\label{eq-cutoff-bc} 
	\end{equation} 
	similarly to the classical case. Such a system is analogous to a flat FLRW model with a potential barrier placed at $\alpha=\alpha_{\text{max}}$. To summarize, we consider the WDW equation for a closed isotropic Universe, where the presence of a self-interaction potential for $\phi$ can play the role of a scattering potential. Even though the solutions of Eq. (\ref{eq-WDW-complete}) are normalizable the Feynman propagator has been found for a set of reparametrized mini-superspace variables that do not have a straightforward interpretation in bouncing scenarios, and we leave this point open for further developments. Furthermore, $V_K(\alpha)$ is negligible near the cosmological singularity. Hence, with the aim of studying the Big Bounce as a scattering process in the framework of RQM, we decided to describe the closed FLRW model with a flat one equipped with the cutoff boundary condition in eq. (\ref{eq-cutoff-bc}). Although such a condition does not take into account the dependence of the turning point from $p_\phi$, it is well suited to describe the Universe whose turning point falls around the fixed $\alpha_{\text{max}}$. Moreover, this choice is supported by the form that the solutions in Eq. (\ref{curve:wdw:bas:sol}) take in the limit $\alpha \to -\infty$. A problem that arises when imposing the condition in eq. (\ref{eq-cutoff-bc}), is that $\hat{p}_\alpha$ in no longer self-adjoint on the half-line $(-\infty, \alpha_{\text{max}}]$. However, by performing numerical computations, it can be observed that the $\Im(\langle \hat{p}_\alpha \rangle) \to 0$ as $\phi \to \pm \infty$, meaning that the momentum operator associated to $\alpha$ can be considered hermitian in that region where the Quantum Bounce is indeed expected to occur, preventing the emergence of possible non-unitarity of the scattering operator. 
	\section{Bounce of the closed isotropic Universe as a quantum transition}
	\begin{figure*}[t]%
		\centering
		\includegraphics[width=1\textwidth]{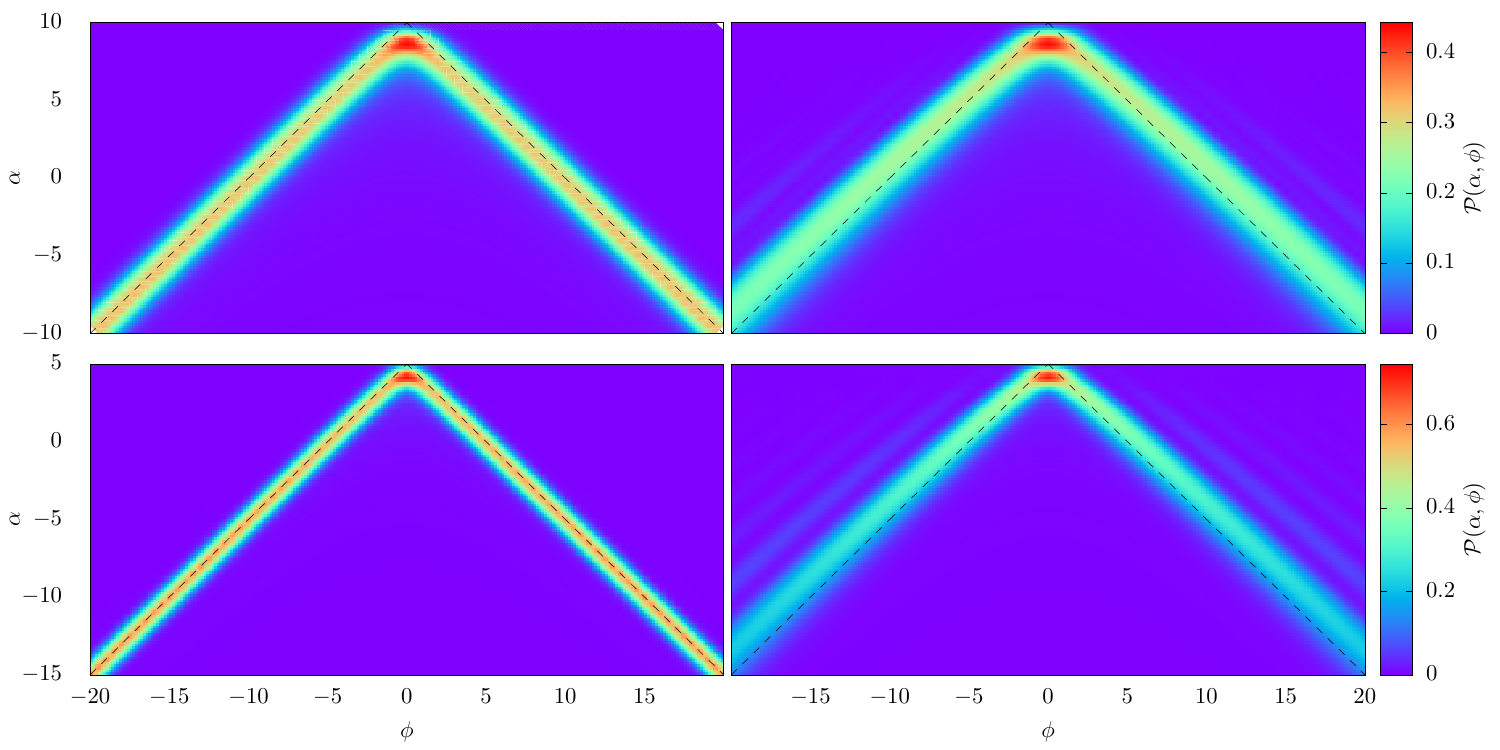}
		\caption{  Plots of the semiclassical evolution computed in the standard QM framework (left) and in PQM one (right) together with the classical evolution law $\alpha_0 (\phi)$ in Eq. (\ref{eq-classical-flat}) (dashed line). We considered positive frequency wave packets with coefficients $A_{\overline{k}, \sigma} (k)\,,\,A^P_{\overline{k}, \sigma}(k)$. On the top row $\overline{k}=0,\, \sigma = 1,\, \alpha_{\text{max}}=10$, $\mu =0.5$, while on the bottom row $\overline{k}=-0.5,\, \sigma = 2,\, \alpha_{\text{max}}=5$, $\mu =0.4$.}
		\label{graph-pdens}
	\end{figure*}%
	In this section, we describe the bounce process of the closed FLRW model as a relativistic quantum transition. The main ingredients of this procedure are a set of ``In''/``Out'' orthonormal states, a time-dependent potential that vanishes at very early/late times, and a Feynman propagator. To fulfill all these requirements, we describe the closed Universe as a flat FLRW model with a cutoff boundary condition. First, we derive the set of basic solutions that represent the ``In''/``Out'' states of the process, providing a physical interpretation of the frequency splitting. Then, we show the Feynman propagator for such a model, and we introduce the potential $U(\phi)$ used in the computation. At last, we compute the probability densities related to the bounce process. This pure quantum treatment of the bounce would be justified if the Universe is described by a de-localized state near the cosmological singularity, and this is not the case of isotropic models. For this reason, we will carry on the discussion and the computations both in the standard quantum mechanical description and in the framework of Polymer Quantum Mechanics (PQM) \cite{Corichi_2007,Corichi_2007_2}, so that the evolution of localized wave packets exhibits a spreading behavior when approaching the cosmological singularity \cite{Giovannetti_2023,Battisti_2008}.  
	\subsection{Solutions of the asymptotically free WDW equation}
	Neglecting $V_K(\alpha)$ from Eq. (\ref{eq-WDW-complete}), the WDW equation we consider is
	\begin{equation}
		\left[ \partial_\alpha^2 - \partial_\phi^2 + C e^{6\alpha} U(\phi) \right] \Psi(\phi, \alpha) = 0\,,
		\label{eq-WDW-approx}
	\end{equation}
	together with the boundary condition in Eq. (\ref{eq-cutoff-bc}). The set of ``In''/``Out'' states is the set of orthonormal solutions of Eq. (\ref{eq-WDW-approx}) in the limit where $U(\phi)\to 0$. In such a regime, Eq. (\ref{eq-WDW-approx}) is completely analogous to a free KG one, and its solutions are plane waves of the form
	\begin{equation}
		f^{L/R}_k (\phi, \alpha) = N_k e^{ik\phi} e^{\pm i p_k \alpha}\,,%
		\label{eq-LR-sol-approx}
	\end{equation}
	where $k \in \mathbb{R}$, $p_k = |k|$ and $N_k$ is a normalization constant. A solution that satisfy Eq. (\ref{eq-cutoff-bc})  must satisfy the continuity condition $\Psi(\phi, \alpha = \alpha_{\text{max}})=0$. For simplicity, let us consider $\alpha_{\text{max}} = 0$ \footnote{Is possible to recover the case of a generic $\alpha_{\text{max}}$ through the transformation $\alpha \to \alpha - \alpha_{\text{max}}$.}. A solution satisfying this continuity condition is in the form $f_k(\phi, \alpha) = f^{L} (\phi, \alpha) - f^{R} (\phi, \alpha)$. It is now possible to perform a frequency splitting of such solutions by choosing the sign of $k$. If we restrict to $k\leq 0$, we find the frequency-separated solutions
	\begin{equation}
		f_k^\pm (\phi, \alpha) = N_k e^{\pm i k \phi} \sin{(k \alpha)}\,.
	\end{equation}
	Imposing the orthonormality relations, like those in Eq. (\ref{eq-orthorel-macd}), we can find the normalization constant $N_k$. Hence the set of ``In/Out'' states of Eq. (\ref{eq-WDW-approx}), when considering a generic $\alpha_{\text{max}}$, is given by the frequency-separated solutions
	\begin{equation}%
		f_k^\pm (\phi, \alpha) = \frac{e^{\pm i k \phi}}{\sqrt{\pi k}} \sin{[k (\alpha - \alpha_{\text{max}})]}\,.%
		\label{eq-qm-wdw-sol-cutoff}
	\end{equation}%
	Let us now apply the formalism of PQM to Eq. (\ref{eq-WDW-approx}). In this particular framework, a generalized coordinate is regarded as discrete, preventing the definition of a proper momentum operator associated with it. Nonetheless, it is possible to define an approximated version of it. Since the effective formulation of LQC is isomorphic to the implementation of PQM to the Mini-superspace variables \cite{Ashtekar_2015_effective, Singh_2005_effective2}, we consider the isotropic degree of freedom $\alpha$ as discrete. Thus, the momentum operator conjugated to $\alpha$ is replaced by its polymer approximated version \cite{Corichi_2007}, yielding the modified WDW equation 
	where 
	\begin{equation}
		\left[-\frac{1}{\mu^2} \sin^2{(\mu\, p_\alpha)} + p_\phi^2 \right] \Psi_P (p_\phi, p_\alpha)= 0\,,%
		\label{eq-poly-wdw-mom}
	\end{equation}
	where $\mu$ is the spacing of the lattice over which $\alpha$ is defined and the limit $U(\phi) \to 0$ has been already considered. The solutions of Eq. (\ref{eq-poly-wdw-mom}) have the form
	\begin{equation}
		g^{L/R}_k (p_\phi, p_\alpha) = \tilde{N}_k\, \delta(p_\phi - k)\, \delta(p_\alpha \pm \tilde{p}_k)\,,
	\end{equation} 
	where $\tilde{p}_k = |\arcsin{(\mu\, k)}|/\mu$ and $k \in [-1/\mu,1/\mu]$. Moving to the ``position'' representation, these solutions are equivalent to the plane waves in Eq. (\ref{eq-LR-sol-approx}) with the modified dispersion relation $\tilde{p}_k$. Thus, the procedure to obtain the set of orthonormal solutions satisfying Eq. (\ref{eq-cutoff-bc}) is completely analogous to that of the standard quantum mechanical framework, yielding
	\begin{equation}
		g^\pm_k (\phi, \alpha) = \frac{e^{\pm i k \phi} \sin[\tilde{p}_k (\alpha - \alpha_{\text{max}})]}{\sqrt{\pi k \sqrt{1-\mu^2 k^2.
		}}}\,,%
		\label{eq-poly-wdw-sol-cutoff}
	\end{equation}  
	where now $k \in [-1/\mu , 0]$. We can immediately notice that, besides the dispersion relation, the standard quantum mechanical case and the polymer one differ by the normalization factor $(1-\mu^2 k^2)^{-1/4}$. { It is worth noting that, the polymer construction, based on the $\sin$ representation, contains a certain degree of ambiguity. In fact, when derived from Loop Quantum Gravity, the specific polymer representation can depend on the Holonomy choice in the underlying theory  \cite{Ashtekar_2011_review}. On the same level, when the polymerization prescription is imposed \emph{a priori} as a regularization of a diffeomorphism invariant formulation \cite{Strocchi:2016kce}, different choices in regularizing the momentum operator are possible. However, when in the Loop Quantum Gravity formulations are adopted the quasi-periodic functions \cite{Bruno_2023,Bruno_2023_2}, as well as, when the prescription on the momentum operator regularization relies on the discretization of the translation operator \cite{Corichi_2007,Corichi_2007_2}, the choice of the sin function appears as the most natural and somehow a privileged representation.} From the solutions in Eq. (\ref{eq-qm-wdw-sol-cutoff}) and in Eq. (\ref{eq-poly-wdw-sol-cutoff}) we can construct localized wave packets and compute their semiclassical evolution. Since, the inner product in Eq. (\ref{eq-KG-inner}) is positive-definite for positive/negative frequency wave function, we can define the probability density
	\begin{equation}
		\mathcal{P}(\alpha, \phi) = \pm i \psi^{\pm*} (\alpha,\phi) \overleftrightarrow{\partial_\phi} \psi^\pm (\alpha, \phi)\,,%
		\label{eq-pdens}
	\end{equation}
	i.e., the probability density of finding the Universe with some value of $\alpha$ at a fixed time $\phi$. Such a quantity is well-defined as long as the condition $\left(\psi^\pm\,,\, \psi^\pm\right)=1$ holds. In Fig. (\ref{graph-pdens}), we show the semiclassical evolution computed from Eq. (\ref{eq-pdens}) for both the standard and the polymer quantum mechanical case {and for different sets of parameters}. We considered the wave packets
    \begin{subequations}%
		\begin{equation}%
			\psi^\pm (\phi, \alpha) = \int_{-\infty}^0 dk\ A_{\overline{k}, \sigma} (k) f_k^\pm (\phi, \alpha)\,,%
			\label{eq-std-wave-packet}%
		\end{equation}%
		\begin{equation}%
			\psi^\pm_P (\phi, \alpha) = \int_{-1/\mu}^0 dk\ A^P_{\overline{k}, \sigma} (k) g_k^\pm (\phi, \alpha)\,,%
			\label{eq-poly-wave-packet}%
		\end{equation}%
		\label{eq-wave-packets}%
	\end{subequations}%
    where $A_{\overline{k}, \sigma} (k)\,,\,A^P_{\overline{k}, \sigma}(k)$, are Gaussian-like coefficients properly normalized over each domain of $k$. We can immediately notice that in both representations of QM, the semiclassical evolution reproduces the classical one. It means that a flat FLRW model, together with the boundary condition in Eq. (\ref{eq-cutoff-bc}), can reproduce the dynamics of the closed isotropic Universe, setting a proper value of $\alpha_{\text{max}}$. The main difference between the standard case and the polymer one lies in the spreading behavior of the polymer wave packet when moving toward the cosmological singularity. Such a feature prevents the notion of semiclassical evolution near $\alpha \to -\infty$, justifying a pure quantum treatment of the Universe in that region. We can now provide an interpretation of the frequency splitting of the solutions in Eq. (\ref{eq-qm-wdw-sol-cutoff},\ref{eq-poly-wdw-sol-cutoff}). Differently from \cite{Giovannetti_2023}, the frequency splitting no longer distinguishes between contraction/expansion phases, since these two branches are linked by the presence of the imposed turning point. Hence, recalling the Feynman interpretation of frequency-separated solutions, the difference between positive/negative frequency wave functions is that one follows the semiclassical evolution forward in time while the other follows it backward in time. To be more precise, positive frequency solutions follow an expanding trajectory from $\phi \to -\infty$ to the turning point, then they undergo a contraction phase towards $\phi \to +\infty$, while negative frequency wave packets follow this trajectory backward in time, i.e. expanding from $\phi \to +\infty$ to the turning point and contracting towards $\phi \to -\infty$. This kind of interpretation pushes even further the analogy between the WDW equation and the KG one, removing the ambiguity in the choice of the sign of $p_\alpha$ that was discussed in \cite{Giovannetti_2023} for the flat FLRW model. 
	\subsection{The effects of the self-interaction potential and the bounce transition}
	\begin{figure*}[t]%
		\centering
		\includegraphics[width=1\textwidth]{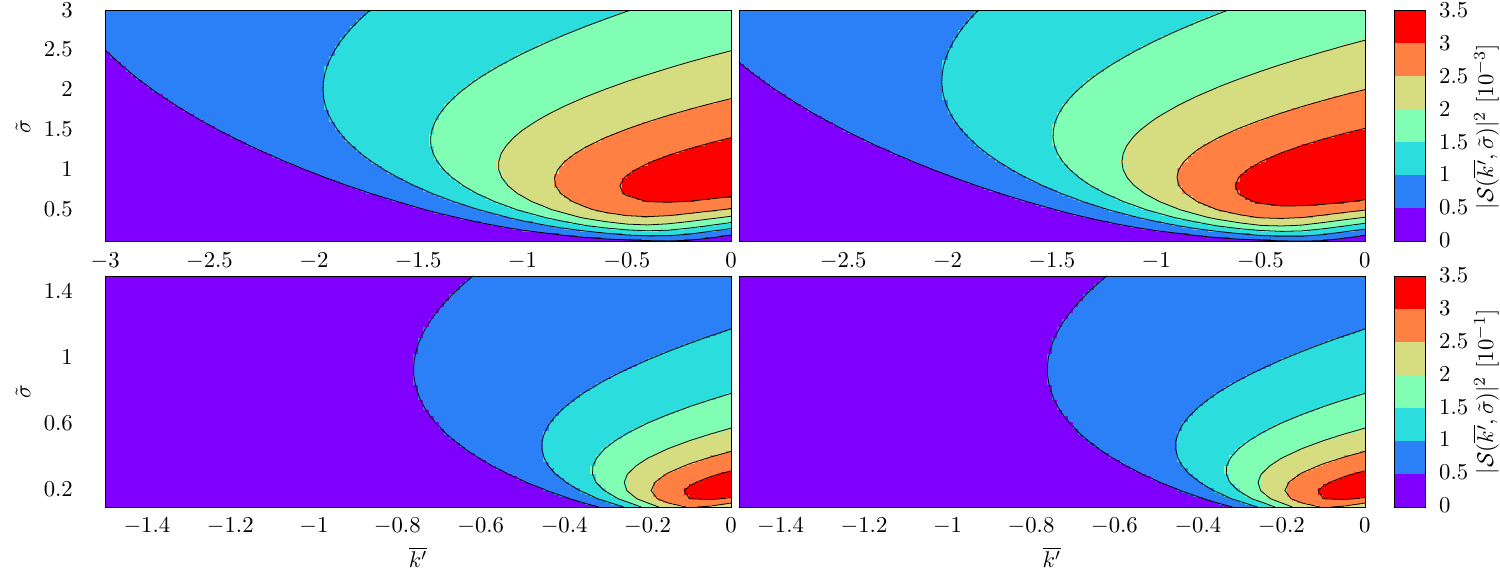}
		\caption{ Plots of the bounce probability density computed in the standard framework (left) and in the polymer framework (right). The dependence from the initial label $\overline{k}$ is integrated out, while $\tilde{\sigma}=\sigma'=\sigma$. On the top row, we considered $\lambda=1/2,\,\gamma = 2,\, \alpha_{\text{max}} = 0$, and the polymer step was set to $\mu =0.1$. On the bottom row, we considered $\lambda=1/2,\,\gamma = 9,\, \alpha_{\text{max}} = 1,\, \mu = 0.02$.}
		\label{graph-amplitudes}
	\end{figure*}%
 
	Until now, we computed the set of asymptotic states from the WDW Eq. (\ref{eq-WDW-approx}). The presence of the time-dependent potential prevents the possibility of a proper frequency splitting so that the Universe is represented by a superposition of positive and negative frequency solutions. The bounce of the Universe can then be described as a quantum transition mediated by the self-interaction potential of the scalar field, as already studied for the Bianchi I Universe in \cite{Giovannetti_2022_bianchi} and for the flat FRLW model in \cite{Giovannetti_2023}. Relying on the interpretation we have given to the solutions of Eq. (\ref{eq-WDW-approx}) and its polymer version in Eq. (\ref{eq-poly-wdw-mom}), the bounce can be described by a transition from a localized negative frequency wave packet to a positive one. Being in a negative frequency state, the Universe propagates from $\phi \to +\infty$, it will contract towards $\alpha \to -\infty$ after reaching the turning point. If the self-interaction potential is turned on near the cosmological singularity, the Universe will be described by a mixed-frequency state. As $U(\phi)$ is turned off, the Universe can perform a transition to a positive frequency localized state, that will again expand and contract towards $\phi \to +\infty$, representing indeed a bounce process. The probability amplitude of this bounce transition can be computed employing the propagator theory of RQM. Eq. (\ref{eq-WDW-approx}) can be equipped with the usual Feynman propagator for the KG equation $\Delta_F (\phi, \phi'; \alpha, \alpha')$. In our case, the propagator must satisfy the boundary condition in Eq. (\ref{eq-cutoff-bc}) as well as propagate forward/backward in time the positive/negative frequency solutions in Eq. (\ref{eq-qm-wdw-sol-cutoff}). Hence, the Feynman propagator we need can be written as 
	\begin{equation}
        \begin{split}
			\Delta_F^{h}(\delta \phi;\delta \alpha)= -i \int_{-\infty}^0 dk\ &[ f^+_k ( \phi, \alpha) f^{+*}_k ( \phi', \alpha') \theta(\phi-\phi')+ \\
            + &f^-_k ( \phi, \alpha) f^{-*}_k ( \phi', \alpha') \theta(\phi'-\phi) ]\,,
        \end{split}
		\label{feynamnprop}
	\end{equation}	
	where the superscript $h$ denotes that we are considering $\alpha \in (-\infty,\alpha_{\text{max}}]$. These considerations are still valid in the polymer framework so that the form of the polymer propagator is completely analogous. Let us now introduce the potential that will be used in the computations. We rely on the ekpyrotic potential studied in \cite{Steinhardt_2002_ekp_pot1,Steinhardt_2002_ekp_pot2,Buchbinder_2007_new_ekp} that is characterized by a global minimum and vanishing in the limit $\phi \to \pm \infty$. Hence, we consider the explicit form
	\begin{equation}
		U(\phi) = -\lambda \frac{\gamma}{\gamma^2 + \phi^2}\,,
		\label{eq-self-pot}
	\end{equation}  
	where $\lambda$ is coupling constant with the dimension of an energy density and $\gamma$ is a width parameter controlling the shape of the potential. Both $\lambda$ and $\gamma$ control the perturbativeness of $U(\phi)$. Such a potential satisfies the properties discussed above. We now have all the elements required to use the propagator scattering theory of RQM. In the light of the interpretation of positive/negative frequency wave packets the bounce can be represented as a quantum transition from a negative frequency solution to a positive one. Let us remark on the fact that in the light of the results from \cite{Wald_1993}, as long as $U(\phi)$ is vanishing asymptotically at least in one direction of the internal time (here it vanishes for both $\phi \to \pm \infty$), it is possible to define a Hilbert space at any time $\phi$. Hence, even if the self-interaction potential mixes the frequency states, it does not alter the Hilbert space in a way that could undermine the unitarity of the scattering operator. At the first order in perturbation theory, the transition amplitude for this process reads as \cite{Bjorken:Drell}
	\begin{equation}
		\mathcal{S} = - i \int d\phi \int d\alpha\ \psi^{+*}(\phi, \alpha) V_S(\phi, \alpha)\psi^{-}(\phi, \alpha)\,,%
		\label{eq-bounce-transition-general} 
	\end{equation}
	where $\psi^\pm(\alpha, \phi)$ are localized wave packets. When considering these wave packets, the probability amplitude of the process depends on the parameters $(\overline{k},\sigma)_{\text{in}}\,,\,(\overline{k'},\sigma')_{\text{out}}$, i.e., $\mathcal{S}=\mathcal{S}(\overline{k'}, \sigma'; \overline{k},\sigma)$. Due to the unitarity of the scattering operator, $|\mathcal{S}|^2$ represents indeed a probability. Thus, from Eq. (\ref{eq-bounce-transition-general}) we can obtain the probability that a negative frequency Universe with a given pair $(\overline{k},\sigma)_{\text{in}}$ exit the interaction with the self-interaction potential being a positive frequency Universe characterized by $(\overline{k'},\sigma')_{\text{out}}$. { By substituting in Eq. (\ref{eq-bounce-transition-general}) the explicit form of the wave packets, together with Eq. (\ref{eq-self-pot}), we can compute the explicit form of the bounce probability amplitude $\mathcal{S}$ for both the standard and the polymer quantum mechanical framework, that in the former case reads as
		\begin{equation}
			\mathcal{S} 
            = \xi \int dk dk' \frac{  A_{\overline{k'},\sigma'}(k') A_{\overline{k},\sigma}(k) e^{\gamma (k+k')} \sqrt{k'k}}{k^4 - 2k^2 (k'^2 - 36) + (k'^2+36)^2 }\,,%
			\label{eq-bounce-transition}%
		\end{equation}
	where $\xi = i 72 \lambda \pi^3 	e^{6\alpha_{\text{max}}}$, while in the latter case
    \begin{equation}
        \mathcal{S}_P = \frac{\xi\mu^2}{8}  \int \frac{dk dk'}{\sqrt{k'k}} \frac{ A^P_{\overline{k'},\sigma'}(k') A^P_{\overline{k},\sigma}(k) e^{\gamma (k+k')}}{[(1-\mu^2 k^2)(1-\mu^2 k'^2)]^{1/4}} [I^\mu_-(k,k') - I^\mu_+(k,k')]\,,
    \end{equation}
    where $I^\mu_\pm(k,k')=[36\mu^2 - (\arcsin(\mu k) \pm \arcsin(\mu k'))^2]^{-1}$. We can notice immediately that in the limits $\lambda \to 0$ and $\gamma \to +\infty$ the probability amplitude vanishes as expected, and $\mathcal{S}_P \to \mathcal{S}$ in the limit $\mu \to 0$.} Moreover, the cutoff $\alpha_{\text{max}}$ contributes only as a global multiplicative factor. The integrals in Eq. (\ref{eq-bounce-transition}) can be computed through numerical integration. In Fig. \ref{graph-amplitudes}, we plot the probabilities {$|\mathcal{S}|^2$ and} $|\mathcal{S}_P|^2$, integrating over all the possible initial labels $\overline{k}$ and varying $\overline{k'}\,,\, \sigma=\sigma'\equiv \tilde{\sigma}$. { Different values of the parameters $\gamma\,,\, \alpha_{\text{max}}\,,$ and $\mu$ have been considered.} From Fig. \ref{graph-amplitudes}, it is clear that when setting a small step $\mu$ the polymer framework yields the same results of the standard one. 
    As we can see, the probabilities are peaked around small values of both $\overline{k'}$ and $\tilde{\sigma}$ while vanishing for both arbitrarily large and small values of them, preventing the emergence of a vanishing energy Universe or with arbitrarily large energy. 
	\section{Discussion of the results}

	In the spirit of \cite{Giovannetti_2022_bianchi,Giovannetti_2023} we described the bounce as a quantum transition between wave packets, that reproduce the expected semiclassical dynamics, one forward and the other backward in time, recovering the Feynman interpretation of positive/negative frequency solutions. Even though these solutions are in principle orthogonal to each other, such a transition is possible due to the presence of a self-interaction potential for the scalar field near the cosmological singularity. The introduction of the cutoff $\alpha_{\text{max}}$ as an approximation of the curvature potential $V_K(\alpha)$ allows us to construct a set of asymptotically free orthonormal states and a proper Feynman propagator. Differently from the standard quantum mechanical case, the pure quantum description of the bounce process becomes necessary when PQM is applied to the isotropic degree of freedom, as the polymer solutions are non-localized near the cosmological singularity, even when $U(\phi)$ is not present. The computations of the bounce probabilities yield approximately the same results both in standard QM and in PQM, i.e., the transition amplitude reaches its maximum in the interval $\overline{k'} \in [-0.5,0]$ and $\tilde{\sigma} \in [0.5,1.5]$. To provide a physical interpretation to these results, it is useful to consider the expectation value of the operator $\hat{p}_\phi$. The wave packets used are characterized by the mean value $\langle \hat{p}_\phi \rangle$ computed with the inner product in Eq. (\ref{eq-KG-inner}). Both for the standard quantum mechanical framework and in the limit $\overline{k} \to 0$ for the polymer one, we have
    \begin{equation}
        \langle \hat{p}_\phi \rangle_\pm = \pm \left( \overline{k} - \frac{\sigma e^{-\overline{k}^2/\sigma^2}}{\pi \erfc{(\overline{k}/\sigma}}) \right) 
    \end{equation}
    From Fig. \ref{graph-amplitudes}, the Universe has the maximum probability of exiting the bounce process in a state characterized by $\langle \hat{p}_\phi \rangle \sim 0$. By comparison with the Hamilton equations in Eqs. (\ref{eq-ham-eqs}), such a characterization is equivalent to the condition $\dot{\alpha} \sim 0$, suggesting the existence of a minimal volume that connects the two phases of the bounce, as derived in \cite{Ashtekar_2007_semiCB_CU,Ashtekar_2011_review}. These are rather different results from those in \cite{Giovannetti_2022_bianchi,Giovannetti_2023}, where the bounce probability was peaked around $\overline{k'} \sim \overline{k}\,,\, \sigma'\sim \sigma$. Such a difference might stem from the use, in those works, of a non-perturbative, or even divergent with respect to the internal time, scattering potential. If one performs all the previous computations using values of $\gamma \ll 1$ that make $U(\phi)$ a non-perturbative potential, we recover this result where the bounce densities reach their maximum in $(\overline{k'},\sigma')_{\text{out}} \sim (\overline{k},\sigma)_{\text{in}}$, supporting our considerations.%
    The possibility to apply the standard formalisms of Relativistic Quantum Mechanics relies on the isomorphism that the Wheeler-DeWitt equation manifests, for the considered model, in comparison to a $1+1$-Klein-Gordon-like equation. The time variable is here identified with the scalar matter field, associated with an ekpyrotic Universe \cite{Steinhardt_2002_ekp_pot1,Steinhardt_2002_ekp_pot2,Buchbinder_2007_new_ekp} and it is a viable (monotonic) physical clock. The possibility of associating the Mini-superspace to a relativistic scenario for quantum mechanics has been first investigated in \cite{Wald_1993, Higuchi_1995}. Actually, it has been shown that the existence of a Hilbert space for the theory comes out even when the potential term of the Wheeler-DeWitt equation is diverging on only one of the two time directions. This result can be directly implemented in the present analysis, simply because our ekpyrotic-like potential vanishes in both $\phi$-directions. This consideration is the starting point to develop the full formalism of a scattering process, as discussed in \cite{Bjorken:Drell}. It is remarkable that, in our formulation, the wave function is always referred to as a single-particle problem (as it takes place in relativistic quantum processes below the threshold of couples creation). Thus, in our study, the possibility of creating or annihilating Universe degrees of freedom is removed, differently from the so-called "third quantization'' approach \cite{Ashtekar_2009_third_1,Ashtekar_2010_third_2}. The important point to be stressed here is that the use of the scattering transition amplitude is a natural consequence of properly implementing the relativistic quantum prescription. In particular, our propagator has exactly the standard meaning of the original Feynman idea: actually the forward and backward in time propagated solutions are here properly identified with the reverse of the time arrow, simply because we pass from expanding to collapsing configurations, at all specular. The resulting transition amplitude, therefore, appears well-grounded on a general quantum prescription and it is natural to attribute a predictivity to its outcoming morphology. A confirmation of such a basically reliable predictivity is also confirmed \emph{a posteriori}, since the maximum for the obtained transition amplitude takes place in correspondence to a quasi-classical bounce picture. This fact stands for its consistency in the cosmological picture, where localized packets are unavoidably involved in dealing with meaningful cosmological states.

	\section{Concluding remarks}
	
	We analyzed the quantum dynamics of the isotropic Universe, in the presence of the self-interacting scalar field in Eq. (\ref{eq-self-pot}), whose potential term was modeled according to an ekpyrotic model. The basic ingredient of our investigation for a quantum bouncing cosmology has been the analogy between the Mini-superspace metric formulation and a Relativistic Quantum Mechanics approach in the physical space-time \cite{Wald_1993, Giovannetti_2022_bianchi}. In the absence of the scalar field potential, thought as a Planckian interaction contribution, the positive and negative frequency solutions are separated, corresponding to the two possible arrows of the time variable, i.e. of the physical clock, provided by the scalar field its-self (in the spirit of a relational time \cite{rovelli_internal_time}). The fundamental novelty of the present formulation with respect to a similar analysis in \cite{Giovannetti_2023} consisted in the really perturbative character of the scattering potential term, having an ekpyrotic morphology, ensuring that it is significantly different from zero only in a finite interval of time and with the amplitude regulated by a coupling constant and a width parameter. The main result obtained above is to be considered the recovering of the quasi-classical condition for a Big-Bounce (i.e. the emergence of a minimal Universe volume configuration) as the most probable value for the transition amplitude. When the parameters of $U(\phi)$ are regulated to deal with a non-perturbative potential, the most probable transition configuration overlaps that one in \cite{Giovannetti_2023}, shedding light on the relevance of dealing with a perturbative scattering potential, when frequency separation is broken.	The results here obtained encourage to extend the proposed picture to more complex systems, not only in cosmology but also including the quantum physics of the gravitational collapse and the so-called ``Black Bounce'' \cite{black-bounce1,black-bounce2}.

\bibliographystyle{ieeetr} 
\bibliography{refs}






\end{document}